\documentclass[]{elsarticle}

\RequirePackage[]{hyperref}
\RequirePackage{amsthm,amsmath,amssymb,bbm}
\usepackage{orcidlink}

\usepackage{rotating}
\usepackage{colortbl}
\usepackage{multirow}
\usepackage{makecell}

\newcommand{\reals}{\mathbb{R}}

\theoremstyle{plain}
\newtheorem{theorem}{Theorem}
\newtheorem{assume}{Assumption}
\newtheorem{lemma}{Lemma}
\newtheorem{cor}{Corollary}[section]

\newtheorem{rmk}{Remark}

\allowdisplaybreaks

\begin{document}
	\begin{frontmatter}
\title{High-dimensional Bayesian Tobit regression for censored response with Horseshoe prior}
		
		\author[aaatienmt]{The Tien Mai\,\orcidlink{0000-0002-3514-9636} }\ead{the.tien.mai@fhi.no}
		
		\affiliation[aaatienmt]{
			organization={
				Norwegian Institute of Public Health}, 
			city={Oslo},
			postcode={0456}, 
			country={Norway}}

		\begin{abstract}
Censored response variables—where outcomes are only partially observed due to known bounds—arise in numerous scientific domains and present serious challenges for regression analysis. The Tobit model, a classical solution for handling left-censoring, has been widely used in economics and beyond. However, with the increasing prevalence of high-dimensional data, where the number of covariates exceeds the sample size, traditional Tobit methods become inadequate. While frequentist approaches for high-dimensional Tobit regression have recently been developed, notably through Lasso-based estimators, the Bayesian literature remains sparse and lacks theoretical guarantees. In this work, we propose a novel Bayesian framework for high-dimensional Tobit regression that addresses both censoring and sparsity. Our method leverages the Horseshoe prior to induce shrinkage and employs a data augmentation strategy to facilitate efficient posterior computation via Gibbs sampling. We establish posterior consistency and derive concentration rates under sparsity, providing the first theoretical results for Bayesian Tobit models in high dimensions. Numerical experiments show that our approach outperforms favorably with the recent Lasso-Tobit method. Our method is implemented in the \texttt{R} package ``\texttt{tobitbayes}", which can be found at \url{https://github.com/tienmt/tobitbayes}.
		\end{abstract}
		
		
\end{frontmatter}

\section{Introduction}
\label{sc_intro}
Censored response variables---where outcomes are only partially observed due to known bounds---are common across diverse fields, for example: environmental science \cite{helsel2011statistics}, economics \cite{mcdonald1980uses}, manufacturing \cite{meeker2021statistical}, and medical research \cite{gandhi2020long}. In such cases, although the exact value of the response is unknown beyond a threshold, it is still known whether the value falls above or below the limit, and the corresponding covariates are observed. This partial information distinguishes censoring from missing data and poses challenges for standard regression methods. Since censoring violates the assumptions underlying ordinary least squares (OLS), applying OLS directly leads to biased and inconsistent estimates \citep{amemiya1984tobit}.

To address this, specialized regression models have been developed to incorporate censoring mechanisms. One of the earliest and most influential is the Tobit model, introduced by \citet{tobin1958estimation}.
Tobin proposed the model to analyze household spending on durable goods, noting that many low-income households reported zero expenditure---a known lower limit rather than a true absence of demand. 
His approach combined elements of probit modeling (to account for the probability of censoring) with linear regression for the uncensored cases, thereby handling the dual nature of the data.
Due to its suitability for analyzing left-censored responses, especially in household surveys and microdata, the Tobit model became a mainstay in applied economic research. 
Over time, it has been extended to accommodate other forms of censoring, including right- and interval-censoring \citep{amemiya1984tobit}, and has found applications beyond its original economic context.

The growing presence of high-dimensional data—where the number of covariates far exceeds the sample size—across a range of disciplines has introduced new analytical challenges, especially in the context of left-censored responses.
Although high-dimensional regression and censored regression have each been the subject of extensive research, relatively few methods have been developed to address both issues simultaneously. Some progress has been made by extending classical estimators: for example, \citet{muller2016censored} and \citet{zhou2016lad} adapted Powell’s least absolute deviation estimator \citep{powell1984least} to the high-dimensional setting. Similarly, the Buckley-James estimator \citep{buckley1979linear} has been extended using regularization techniques by \citet{johnson2009lasso}, \citet{li2014dantzig}, and \citet{soret2018lasso}. Most recently, \citet{jacobson2024high} proposed a direct high-dimensional extension of the Tobit model using the Lasso, offering a unified approach for sparse estimation in the presence of left-censoring.

While the Bayesian treatment of Tobit models has historically received less attention than frequentist approaches, it dates back to the seminal work of \citet{chib1992bayes}. 
Since then, a variety of Bayesian extensions have been proposed. For instance, \citet{yu2007bayesian,zhao2015bayesian,alhamzawi2018bayesian,alhusseini2018bayesian,alhamzawi2015bayesian,kobayashi2017bayesian} proposed  Bayesian analysis of  Tobit quantile regression models; 
\citet{alhamzawi2016bayesian,abbas2019bayesian,alhamzawi2020new} developed Bayesian elastic net for Tobit quantile regression, Bayesian model selection is considered in \cite{ji2012model}. More recently, methodological innovations have continued to emerge: \citet{basson2023variational} proposed a scalable variational inference framework for Tobit models, and \citet{o2024type} introduced a Bayesian additive regression trees (BART) approach tailored to censored data. For a comprehensive overview of recent computational advancements, see the review by \citet{anceschi2023bayesian}.

Despite this growing literature, a notable gap remains: there are currently no theoretical guarantees for Bayesian Tobit regression methods in high-dimensional regimes. In contrast, the past decade has witnessed significant theoretical progress in high-dimensional Bayesian inference, particularly for linear models \citep{castillo2015bayesian} and generalized linear models (GLMs) \citep{jeong2021posterior}. However, these theoretical frameworks cannot be directly extended to the Tobit model. The presence of censoring introduces structural discontinuities and alters the likelihood in ways that violate key assumptions underlying standard Bayesian theory for linear and generalized linear models. As a result, establishing posterior consistency or contraction rates for high-dimensional Bayesian Tobit models presents unique challenges that have yet to be  addressed.

To address this gap, we propose a novel Bayesian framework for sparse Tobit regression that employs the fractional posterior---a generalization of the standard Bayesian posterior obtained by raising the likelihood to a fractional power \citep{bhattacharya2016bayesian}.
This approach has not previously been considered in the context of censored regression and offers multiple advantages.  
From a theoretical standpoint, the fractional posterior also facilitates sharper concentration results and is well-suited for high-dimensional analysis \citep{alquier2020concentration}.
Our contribution builds on and extends recent developments in generalized Bayesian inference \citep{bissiri2013general}. 
It also aligns with contemporary trends in scalable inference, such as variational approximations for fractional posteriors \citep{Knoblauch}, and data-driven tuning strategies inspired by empirical Bayes \citep{martin2020empirical}. 
By placing our methodology within this broader framework, we highlight its versatility for both theoretical exploration and practical deployment.

Theoretically, we derive posterior consistency and concentration rates for the proposed fractional posterior under sparsity assumptions. These results represent, to our knowledge, the first such guarantees for Bayesian Tobit models in high dimensions, complementing recent frequentist advances such as those by \citet{jacobson2024high}. By doing so, we provide a theoretically rigorous and practically robust alternative to existing regularization methods, with implications for a wide range of applications involving censored and sparse data.

To address the computational challenge of sparsity in high-dimensional censored regression settings, our Bayesian Tobit regression model incorporates the use of the Horseshoe prior \citep{carvalho2010horseshoe}. To the best of our knowledge, this marks the first formal integration of the Horseshoe prior into the Tobit framework, bridging two important strands of modern statistical modeling: censoring mechanisms and sparsity-inducing priors. The Horseshoe prior has gained prominence for its ability to perform aggressive shrinkage of noise coefficients while retaining signals, making it highly suitable for sparse, high-dimensional data \citep{makalic2015simple,bhadra2015horseshoe_,piironen2017sparsity}. We exploit its hierarchical scale-mixture structure as in \cite{makalic2015simple} to construct a fully Bayesian model that captures both uncertainty and sparsity.

A key technical innovation lies in our tailored data augmentation scheme, which enables efficient posterior computation despite the nonstandard censoring mechanism. By deriving closed-form full conditional distributions for all model parameters, we develop a novel Gibbs sampler that ensures computationally tractable inference while maintaining exact Bayesian validity. This sampler leverages both the latent variable formulation of the Tobit model and the Horseshoe prior’s latent scale parameters, resulting in a modular and scalable inference procedure.

Our method is validated through extensive simulation studies across a range of high-dimensional scenarios, including those with varying degrees of sparsity and censoring. Results consistently demonstrate superior predictive accuracy and estimation performance relative to the Lasso-based Tobit approach of \citet{jacobson2024high}, particularly in regimes where sparsity is pronounced. Additionally, we apply our method to a real dataset to showcase its practical utility and interpretability.

The remainder of the paper is organized as follows. Section \ref{sc_model_method} introduces the sparse Tobit regression model along with our Bayesian formulation using the Horseshoe prior, and presents the associated theoretical results. In Section \ref{sc_gibbsampler}, we detail the Gibbs sampling algorithm developed for posterior inference. Section \ref{sc_numerical} reports the results of simulation studies and a real data application, including comparisons with the Lasso-based Tobit method. We conclude with a discussion in Section \ref{sc_conclusion}. Technical proofs are provided in the Appendix (\ref{sc_proofs}).

\section{Bayesian sparse Tobit regression}
\label{sc_model_method}
\subsection{Sparse Tobit model for censored response}
Suppose that we observe a set of predictors, $x_1, \ldots, x_p$, and a response $y \geq c$ where $c$ is a known lower limit (for example, $c = 0$).
In Tobit regression we assume that there exists a latent response variable $y^*$ such that 
$$
y = \max\{y^*, c\}
$$
and that $y^*$ comes from a linear model
$$
y^* =  X^\top \beta + \epsilon
,
$$
where $X = (x_1, \ldots, x_p)^\top  \in \reals^{p}$, $\beta = ( \beta_1, \ldots, \beta_p) \in \reals^{p} $, and $\epsilon \sim N(0, \sigma^2)$. Here, we assume that $ \sigma $ is known and further assume that $ \sigma =1 $.
In the following developments we assume that $c=0$ without loss of generality.

Let \( P_\beta \) denote the joint probability distribution of \( (y, X) \) under the parameter \( \beta \). Given \( n \) i.i.d. observations \( \{(y_i, X_i)\}_{i=1}^n  \) of the pair \( (y, X) \), 
we define \( d_i = \mathbb{I}_{\{y_i > 0\}} \) as an indicator of whether observation \( i \) is uncensored. Using the latent-variable formulation and the properties of the normal distribution, we can derive the likelihood function for the above Tobit regression as
$$
L_n(\beta) 
= 
\prod_{i=1}^{n} p_{\beta} (y_i,X_i)
= 
\prod_{i=1}^{n} \left[ 
\frac{1}{\sqrt{2\pi}}
\exp\left\{-\frac{ (y_i - X_i^\top  \beta)^2 }{2} \right\} \right]^{d_i} 
\left[ \Phi ( - X_i^\top  \beta ) \right]^{1-d_i} 
,
$$
where $\Phi(\cdot)$ denote the standard normal CDF.

Let $ \beta_0 $ be the true underlying parameter of our model. 
In this study, we examine a high-dimensional sparse scenario, thus assuming that $ s^* < n < d $, where $ s^*:= \|\beta_0 \|_0 $.

\subsection{Method}

We consider a sparse (generalized) Bayesian approach for high-dimensional Tobit regression. More specifically, for $\alpha \in (0,1) $ and a sparsity prior $  \pi (\beta) $ given in \eqref{eq:HS}, we study the following (fractional) posterior for $ \beta$
\begin{align}
    \label{eq_frac_posterior}
    \pi_{n, \alpha } (\beta)
\propto
L_n(\beta)^{\alpha} \pi (\beta), 
\end{align}
using the notation of \cite{bhattacharya2016bayesian}. 
Setting $\alpha = 1$ yields the standard Bayesian posterior. By allowing $\alpha < 1$, we temper the likelihood slightly, which helps guard against model misspecification \citep{alquier2020concentration}, while choosing values close to 1 (e.g., $\alpha = 0.99$ in practice) ensures the inference remains close to that of conventional Bayesian analysis \citep{martin2017empirical}.

The horseshoe prior, \cite{carvalho2010horseshoe}, is independently specified on the elements of $ \beta $ as 
\begin{equation}
\label{eq:HS}
\begin{aligned}
\beta_{j} \mid \lambda_{j},\tau 
& \sim  
\mathcal{N} (0,\lambda_{j}^2 \tau^2), 
\\
\lambda_{j} 
& \sim \mbox{Cau}_+(0, 1), 
\\ 
\tau 
& \sim \mbox{Cau}_+(0, 1)
,
\end{aligned}
\end{equation}
for $ 1 \leq j \leq p$, where $\mbox{Cau}_+(0, 1)$ denotes the truncated standard half-Cauchy distribution with density proportional to $(1+u^2)^{-1} \mathbbm{1}_{(0, \infty)}(u)$. We shall denote this prior induced by the hierarchy in \eqref{eq:HS} by $\pi_{HS} $.

\subsection{Theoretical result}

Let $\alpha\in(0,1)$ and $P,R$ be two probability measures. Let $\mu$ be any measure such that $P\ll \mu$ and $R\ll \mu$. The $\alpha$-R\'enyi divergence  between two probability distributions $P$ and $R$ is  defined by
\begin{align*}
D_{\alpha}(P,R)  =
\frac{1}{\alpha-1} \log \int \left(\frac{{\rm d}P}{{\rm d}\mu}\right)^\alpha \left(\frac{{\rm d}R}{{\rm d}\mu}\right)^{1-\alpha} {\rm d}\mu  \text{,}
\end{align*}
and the Kullback-Leibler divergence is defined by
$
\mathcal{K}(P,R)  = 
\int \log \left(\frac{{\rm d}P}{{\rm d}R} \right){\rm d}P $  if  $ P \ll R
$, and  $
+ \infty$ otherwise.

Hereafter, we formulate some required conditions for our theoretical analysis.

\begin{assume}
\label{assum_grow_p}
    It is assumed that  \( p\) can grow at most as \( \exp(n^b) \) for some \( b < 1 \).
\end{assume}

\begin{assume}
\label{assum_beta0_bounded}
    Assume that there exists a positive constant $C_1 $ such that $ \|\beta_0\|_\infty \leq C_1  $.
\end{assume}

\begin{assume}
\label{asmum_lip_alquier}
We assume that there is a measurable real function $ h(\cdot) $ such that for all $ \beta $ satisfies $
\|\beta -\beta_0\|_2<\delta  $ with $ \delta = [ s^*\log (p/s^*)/n ]^{1/2} $, we have
\[ 
\vert\log p_{\beta_0} (X)
-
\log p_{\beta}(X) \vert 
\leq 
h(X) \Vert\beta-\beta_0 \Vert_2^2
.
\]
Furthermore we assume that $\mathbb{E} h(X)=:B_1,\, \mathbb{E} h^2(X)=:B_2 < \infty $.
\end{assume}

It should be noted that Assumption \ref{asmum_lip_alquier} has been used before in \cite{alquier2020concentration} to study variational inference methods. This assumption is satisfied for example when using logistic regression. However, in our Tobit regression, we need this assumption to control the log-likelihood ratio not to be explored. More specially, similar assumption has been used for penalized Tobit regression in \cite{jacobson2024high}. In Generalized linear model, similar kinds of assumption have also been made as documented in \cite{van2008high,rigollet2012kullback,abramovich2016model}

\begin{theorem}
\label{theorem_result_dis_expectation}
	For any $\alpha\in(0,1)$, 
assume that Assumption \ref{assum_grow_p}, \ref{assum_beta0_bounded} and \ref{asmum_lip_alquier} hold. We have that
\begin{equation}
\label{eq_consistency}
  \mathbb{E} 
\left[ \int D_{\alpha}(P_{\beta},P_{\beta_0}) \pi_{n,\alpha}({\rm d}\beta ) \right]
\leq \frac{1+\alpha}{1-\alpha}\varepsilon_n
	,
\end{equation}
    \begin{equation}
 \mathbb{P}\left[
\int D_{\alpha}(P_{\beta},P_{\beta_0}) \pi_{n,\alpha}({\rm d}\beta )  
\leq 
\frac{2(\alpha+1)}{1-\alpha} \varepsilon_n\right] 
  \geq 
1-\frac{2}{n\varepsilon_n}
,
\label{eq_concentration}
\end{equation}
 where
 $
 \varepsilon_n
 =
K   s^* \log \left( p /s^*\right) / n
 $, for some numerical constant $ K>0 $ depending only on $C_1, B_1, B_2 $.
  \end{theorem}

We remind that all technical proofs are given in \ref{sc_proofs}.

Theorem \ref{theorem_result_dis_expectation} establishes key theoretical guarantees for our method where our main technical point is based on a general results given in \cite{bhattacharya2016bayesian,alquier2020concentration}. Some recent applications of generalized Bayesian method are explored in \cite{mai2024concentration,mai2024gbayslogistic,mai2024properties,mai2025concentration}.
More particularly,
\begin{itemize}
    \item inequality \eqref{eq_consistency} demonstrates that the fractional posterior achieves consistency with respect to the $\alpha$-R\'enyi divergence, meaning that as the sample size increases (under Assumption \ref{assum_grow_p}), 
the posterior distribution for $\beta$ concentrates increasingly tightly around the true parameter $\beta_0$. 

\item  Inequality \eqref{eq_concentration} provides a sharper result, quantifying this concentration at a specific rate $\varepsilon_n$. This establishes not only that the method learns the truth asymptotically, but also how quickly it does so---offering meaningful insight into the efficiency of the fractional posterior in high-dimensional, sparse settings. 
More specifically, the probability bound in \eqref{eq_concentration} is at least $ 1- 2/[K   s^* \log \left( p /s^*\right)] $ which remains nontrivial even in high-dimensional settings where 
$p>n$. 
\end{itemize}
This ensures that our theoretical guarantees are meaningful and applicable in high-dimensional sparse Tobit regression.

By leveraging results from \cite{van2014renyi} regarding the connections between the Hellinger distance, the total variation and the $\alpha$-R\'enyi divergence, we can obtain the following concentration results: 
\begin{align*}
    \mathbb{P}\left[
\int H^2 (P_{\beta},P_{\beta_0}) \pi_{n,\alpha}({\rm d}\beta )  
\leq 
K_\alpha
\varepsilon_n\right] 
 & \geq 
1-\frac{2}{n\varepsilon_n}
,
\\
\mathbb{P}\left[
\int d^{2}_{TV} (P_{\beta},P_{\beta_0}) \pi_{n,\alpha}({\rm d}\beta )  
\leq 
K_\alpha \varepsilon_n\right] 
 & \geq 
1-\frac{2}{n\varepsilon_n}
,
\end{align*}
where $d_{TV}$ being the total variation distance and $ H^2 $ being squared Hellinger distance and $ K_\alpha $ being some postive constant depending only $ \alpha $.

Under additional assumption, we can derive concentration results in term of Euclidean distance. This is given in the following corollary.

\begin{cor}
\label{cor_1}
Assume that Theorem \ref{theorem_result_dis_expectation} holds true and assume that there is a constant $ B_3 >0 $ such that for all $ \beta $ satisfies $
\|\beta -\beta_0\|_2< [ s^*\log (p/s^*)/n ]^{1/2} $, we have
$
\vert\log p_{\beta_0} (X)
-
\log p_{\beta}(X) \vert 
\geq 
B_3 \Vert X^\top \! \beta- X^\top \! \beta_0 \Vert_2^2
.
$
Then, We have that
\begin{equation}
  \mathbb{E} 
\left[ \int \Vert X^\top \! (\beta-  \beta_0) \Vert_2^2 \pi_{n,\alpha}({\rm d}\beta ) \right]
\leq \frac{1+\alpha}{(1-\alpha) B_3 } \varepsilon_n
	,
\end{equation}
\begin{equation}
 \mathbb{P}\left[
\int \Vert X^\top \! (\beta-  \beta_0) \Vert_2^2 
\pi_{n,\alpha}({\rm d}\beta )  
\leq 
\frac{2(\alpha+1)}{ (1-\alpha) B_3 } 
\varepsilon_n\right] 
  \geq 
1-\frac{2}{n\varepsilon_n}
.
\end{equation}
\end{cor}

\begin{rmk}
In \cite{jacobson2024high}, the authors established that the Lasso estimator in 
Tobit regression achieves an $\ell_2$-error rate of order $s^* \log(p) / n$. When a restricted eigenvalue condition is assumed, as in  \cite{jacobson2024high}, from our Corollary \ref{cor_1}
the squared $\ell_2$-error $\|\beta - \beta_0\|_2^2$ can be bounded by $s^* \log(p / s^*) / n$. This matches the minimax-optimal rate for sparse linear models, as shown in \cite{bellec2018slope}.
\end{rmk}

\section{A Gibbs sampling implementation}
\label{sc_gibbsampler}
We formulate our model within a latent variable framework (data augmentation \cite{chib1992bayes}), as follows:
\[
\begin{aligned}
z_i \mid \beta, \sigma^2 &\sim \mathcal{N}(x_i^\top \beta, \sigma^2) ,
\quad
i = 1, \ldots , n ,
\\
y_i &= \max(0, z_i) \quad \text{(left-censored at 0)}
.
\end{aligned}
\]
For simplicity, we assume the censoring occurs at 0, though it is straightforward to generalize this to an arbitrary censoring level $c$.
Using the re-parametrization trick as in \cite{makalic2015simple}, the Horseshoe prior for coefficients is written as:
\[
\begin{aligned}
\beta_j \mid \lambda_j, \tau &\sim \mathcal{N}(0, \tau^2 \lambda_j^2),
\quad j = 1, \ldots , p,  
\\
\lambda_j^2 \mid \nu_j &\sim \text{Inv-Gamma}\left(\frac{1}{2}, \frac{1}{\nu_j} \right), 
\quad 
\nu_j \sim \text{Inv-Gamma}\left(\frac{1}{2}, 1 \right) ,
\\
\tau^2 \mid \xi &\sim \text{Inv-Gamma}\left(\frac{1}{2}, \frac{1}{\xi} \right), 
\quad 
\xi \sim \text{Inv-Gamma}\left(\frac{1}{2}, 1 \right)
.
\end{aligned}
\]
Prior for error variance:
\[
\sigma^2 \sim \text{Inv-Gamma}(a_0, b_0)
.
\]

Our goal is to sample from posterior given the joint posterior:
\[
\pi ( z, \beta, \lambda, \nu, \tau, \xi, \sigma^2 \mid y, X) 
\propto 
\pi (y \mid z) 
\pi (z \mid \beta, \sigma^2) 
\pi (\beta \mid \lambda, \tau) 
\pi (\lambda \mid \nu) 
\pi (\nu) 
\pi (\tau \mid \xi) 
\pi (\xi) \pi (\sigma^2)
.
\]
We derive each full conditional by conditioning on all other parameters and sampling from the resulting distribution.
Hereafter, we provide a step-by-step Gibbs Sampler with full posterior conditioning:
\begin{itemize}
\item[Step 1:] Sample \( z_i \mid y_i, \beta, \sigma^2 \)

For each \( i \):
\\
- If \( y_i > 0 \), we know \( z_i = y_i \).
\\
- If \( y_i = 0 \), then \( z_i \le 0 \), and the likelihood is:
\[
\pi (z_i \mid \beta, \sigma^2, y_i=0) \propto \mathcal{N}(z_i \mid x_i^\top \beta, \sigma^2) \cdot \mathbb{I}(z_i \le 0)
.
\]
Thus, we have the full conditional:
\[
z_i \mid y_i, \beta, \sigma^2
\sim 
\mathcal{N}(x_i^\top \beta, \sigma^2) \text{ truncated to } (-\infty, 0],
\quad 
\text{if } y_i = 0
.
\]

\item[Step 2:] Sample \( \beta \mid z, \lambda, \tau, \sigma^2 \)

- Likelihood: \( z \mid \beta \sim \mathcal{N}(X\beta, \sigma^2 I) \).
\\
- Prior: \( \beta \sim \mathcal{N}(0, \tau^2 \Lambda) \), where \( \Lambda = \text{diag}(\lambda_1^2, \dots, \lambda_p^2) \).
\\
   Using Bayes' rule:
\[
\begin{aligned}
p(\beta \mid z, \lambda, \tau, \sigma^2 )
&\propto 
p(z \mid \beta, \sigma^2) \cdot p(\beta \mid \lambda, \tau) \\
&\propto \exp\left(-\frac{1}{2\sigma^2} \|z - X\beta\|^2\right) \cdot \exp\left(-\frac{1}{2} \beta^\top D^{-1} \beta \right)
,
\end{aligned}
\]
where \( D = \tau^2 \Lambda \). Thus, we obtain the full conditional as:
\[
\beta \mid z, \lambda, \tau, \sigma^2 
\sim \mathcal{N}(\mu_\beta, \Sigma_\beta), 
$$
$$
\text{ where}
\quad
\Sigma_\beta = \left(\frac{1}{\sigma^2} X^\top X + D^{-1} \right)^{-1}, 
\quad
\mu_\beta = \Sigma_\beta \cdot \frac{1}{\sigma^2} X^\top z
.
\]

\item[Step 3:] 
Sample \( \lambda_j^2 \mid \beta_j, \tau, \nu_j \)

Full conditional:
\[
p(\lambda_j^2 \mid \beta_j, \tau, \nu_j ) 
\propto 
\mathcal{N}(\beta_j \mid 0, \tau^2 \lambda_j^2) 
\cdot 
\text{Inv-Gamma}\left(\lambda_j^2 \mid \frac{1}{2}, \frac{1}{\nu_j} \right)
.
\]
This is conjugate and gives:
\[
\lambda_j^2 \mid \beta_j, \tau, \nu_j
\sim 
\text{Inv-Gamma}\left(1, \frac{1}{\nu_j} + \frac{\beta_j^2}{2\tau^2} \right)
.
\]

\item[Step 4:] 
Sample \( \nu_j \mid \lambda_j^2 \)

From the inverse-gamma prior and the previous step:
\[
\nu_j  \mid \lambda_j^2  \sim \text{Inv-Gamma} \left( 1, 1 + \frac{1}{\lambda_j^2}\right)
.
\]

\item[Step 5:]
Sample \( \tau^2 \mid \beta, \lambda, \xi \)

We have that
\[
p(\tau^2 \mid \beta, \lambda, \xi ) 
\propto 
\prod_{j=1}^p \mathcal{N}(\beta_j \mid 0, \tau^2 \lambda_j^2) 
\cdot 
\text{Inv-Gamma} \left(\tau^2 \mid \frac{1}{2}, \frac{1}{\xi} \right)
.
\]
This leads to:
\[
\tau^2 \mid \beta, \lambda, \xi 
\sim 
\text{Inv-Gamma}\left(\frac{p+1}{2}, \frac{1}{\xi} + \frac{1}{2} \sum_{j=1}^p \frac{\beta_j^2}{\lambda_j^2} \right)
.
\]

\item[Step 6:] 
Sample \( \xi \mid \tau^2 \)
\[
\xi \mid \tau^2
\sim \text{Inv-Gamma} \left( 1, 1 + \frac{1}{\tau^2} \right)
.
\]

\item[Step 7:] 
Sample \( \sigma^2 \mid z, \beta \)

We have that
\[
p(\sigma^2 \mid z, \beta) \propto \mathcal{N}(z \mid X\beta, \sigma^2 I) \cdot \text{Inv-Gamma}(a_0, b_0)
.
\]
This gives:
\[
\sigma^2  \mid z, \beta  \sim \text{Inv-Gamma}\left(a_0 + \frac{n}{2}, \; b_0 + \frac{1}{2} \sum_{i=1}^n (z_i - x_i^\top \beta)^2 \right)
.
\]
\end{itemize}

In general, the Gibbs sampling procedure begins with appropriate initial values for all model parameters and latent variables. The sampler then proceeds by iteratively cycling through the full sequence of conditional distributions specified in Steps 1 through Step 7. At each iteration, new values are drawn for the latent variable $z$, the regression coefficients $\beta$, the local shrinkage parameters $\lambda = (\lambda_1, \dots, \lambda_p)$, their associated auxiliary variables $\nu = (\nu_1, \dots, \nu_p)$, the global shrinkage parameter $\tau$, its corresponding auxiliary variable $\xi$, and the residual variance $\sigma^2$. This iterative procedure generates samples from the joint posterior distribution of all unknown quantities in the model. After a suitable burn-in period, the retained samples can be used to approximate posterior summaries and conduct inference on model parameters.

Our method is accessible via the \texttt{R} package ``\texttt{tobitbayes}", which can be found at \url{https://github.com/tienmt/tobitbayes}.

\section{Numerical studies}
\label{sc_numerical}
\subsection{Setup}
We generate an uncensored response from a linear model  $y_i^* = X_i^\top \beta_0 + \epsilon_i$,
where $ X_i \sim N(0, \Sigma)$ and $\epsilon_i \sim N(0, 1)$, and left-censor it to create $y_i$ as follows
$$
y_i = \max\{y_i^*, 0 \}
.
$$  
We run simulations with each of the following covariance structures for the predictors: independent $ \Sigma = \mathbb{I}_p $ and
correlation
$(\Sigma)_{ij} = \rho_X^{|i-j|}$ for all $i,j$.
We consider three settings:
\begin{itemize}
    \item a small scale data with $ n =80, p =100 $;
    \item a medium scale with $ n =200, p =300 $; and
    \item a larger scale with $ n = 500, p = 1000 $.
\end{itemize}
In these contexts, the sparsity \( s^* \) varies between \( s^* = 50 \) and \( s^* = 5 \), with the latter case corresponding to a highly sparse model. 
The first haft of non-zero coefficients of \( \beta_0 \) are set to 1 and the second haft are set to $ -1 $.

We consider the following error metrics to quantify the estimation accuracy of our proposed method against the Lasso Tobit in \cite{jacobson2024high}: 
$$
\ell_2(\beta_0) : = p^{-1} \| \widehat{\beta} -\beta_0 \|_2^2 ,
$$
$$
\ell_2( X^\top \!\! \beta_0)
: = 
n^{-1} \| X^\top \! \widehat{\beta} - X^\top \! \! \beta_0 \|_2^2 ,
$$
in which $ \widehat{\beta} $ is the considered methods, and the following error metrics to quantify the prediction accuracy
\begin{align*}
  \ell_2(Y)
  &: = n^{-1} \sum_{i=1}^n \left( y_i  -\max\{X_i^\top\widehat{\beta}, 0 \}  \right)^2 ,
\\
\ell_2(Y_{\rm test})
& := 
n_{\rm test}^{-1} \sum_{i=1}^{n_{\rm test}} \left( y_{{\rm test},i}  -\max\{X_{{\rm test},i}^\top\widehat{\beta}, 0 \}  \right)^2  
.
\end{align*}
Here, $ y_{{\rm test}} $ and $ X_{{\rm test}} $ are testing data generated as $y $ (using $\beta_0 $) and we take $ n_{\rm test} =30 $ for all setting. 
We run Gibbs sampler for 1200 steps and discard the first 500 steps as burn-in and $ \alpha $ is fixed to 0.99.
Our method is denoted as ``Horseshoe".
For Lasso from \cite{jacobson2024high}, we use default options and use 3 folds cross-validation for selecting the best tuning parameter.  
For each simulation setting, we generate independent $100$ datasets and report the average results together with its standard deviations. We note that simulations comparision between Lasso for Tobit with different frequentist methods have been done in \cite{jacobson2024high}.

\begin{table}[!ht]
\centering
	\caption{Simulation results for $ p = 100, n = 80 $.}
	\begin{tabular}{  l ccc c  }
		\hline \hline
Method  & $ 10 \times \ell_2(\beta_0) $ 
& $ \ell_2( X^\top\!\! \beta_0) $ 
& $ \ell_2(Y)$ 
& $ \ell_2(Y_{\rm test}) $ 
		\\
		\hline
\multicolumn{5}{c }{ $ s^* = 10 $ }
\\ 
\hline
Horseshoe 
& 0.14 (0.10) & 0.89 (0.39) & 0.09 (0.08) & 1.12 (0.68)
		\\
Lasso
&  0.23 (0.11) & 1.40 (0.61) & 0.43 (0.33) & 1.52 (0.84)
		\\
		\hline
\multicolumn{5}{c }{ $ s^* = 10,  \rho_X = 0.5 $ }
\\ 
\hline
Horseshoe 
& 0.26 (0.13) & 1.04 (0.44) & 0.11 (0.07) & 1.12 (0.56)
		\\
Lasso 
&  0.19 (0.09) & 1.22 (0.56) & 0.46 (0.25) & 1.17 (0.54)
\\
		\hline
\multicolumn{5}{c }{ $ s^* = 50 $ }
\\ 
\hline
Horseshoe 
& 3.26 (0.51) & 10.02 (3.11) & 1.11 (1.62) & 13.4  (5.05)
		\\
Lasso
& 3.53 (0.58) & 15.04  (8.28) & 4.13 (4.69) & 14.7  (5.92)
		\\
		\hline
\multicolumn{5}{c }{ $ s^* = 50,  \rho_X = 0.5  $ }
\\ 
\hline
Horseshoe 
& 3.24 (0.89) & 11.66  (4.02) & 0.03 (0.03) & 14.05  (7.60)
		\\
Lasso
& 3.08 (0.49) & 17.05  (7.09) & 2.41 (1.82) & 16.94  (7.63)
		\\
		\hline
		\hline	
\end{tabular}
\label{tb_p100}
\end{table}

\subsection{Simulation results}

In general, the results given in Tables \ref{tb_p100}, \ref{tb_p300},  \ref{tb_p1000n500} and \ref{tb_largesample} show that the Horseshoe method tends to outperform the Lasso method across various scenarios. However, when there is correlation among the covariates, the estimation error of the Horseshoe method is slightly higher, though its prediction error remains highly competitive.

The results shown in Tables \ref{tb_p100}, \ref{tb_p300}, and \ref{tb_p1000n500} clearly demonstrate that both estimation and prediction errors increase as the number of non-zero components in the underlying parameter vector rises. This supports our theoretical findings.

As both $p$ and $n$ increase, from Table \ref{tb_p100} to Table \ref{tb_p1000n500}, we observe that both methods become more stable. In most cases, the Horseshoe method outperforms Lasso in terms of both estimation and prediction errors. This difference is particularly noticeable in the case where $p = 1000$ and $n = 500$ (Table \ref{tb_p1000n500}) for $s^* = 50$, where the Horseshoe method outperforms Lasso by about 10 times.

Regarding correlation in the covariate matrix, in some cases, such as in Tables \ref{tb_p100}  and Table \ref{tb_p300} with $ s^* = 50 $, Lasso performs slightly better than Horseshoe. However, this advantage disappears in higher-dimensional settings, such as in Table \ref{tb_p1000n500}, where the Horseshoe method consistently outperforms Lasso.

\begin{table}[!h]
\centering
	\caption{Simulation results for $ p = 300, n = 200 $.}
	\begin{tabular}{  l ccc c  }
		\hline \hline
Method  & $ 100 \times \ell_2(\beta_0)  $ 
& $ \ell_2( X^\top\!\! \beta_0) $ 
& $ \ell_2(Y)$ 
& $ \ell_2(Y_{\rm test}) $ 
		\\
		\hline
\multicolumn{5}{c }{ $ s^* = 10 $ }
\\ 
\hline
Horseshoe 
& 0.08 (0.03) & 0.29 (0.09) & 0.16 (0.05) & 0.57 (0.24)
		\\
Lasso
& 0.27 (0.09) & 0.65 (0.22) & 0.45 (0.16) & 0.88 (0.43)
		\\
		\hline
\multicolumn{5}{c }{ $ s^* = 10,  \rho_X = 0.5 $ }
\\ 
\hline
Horseshoe 
& 0.11 (0.04) & 0.28 (0.09) & 0.17 (0.05) & 0.64 (0.27)
		\\
Lasso 
& 0.23 (0.09) & 0.57 (0.20) & 0.46 (0.16) & 0.81 (0.35)
\\
		\hline
\multicolumn{5}{c }{ $ s^* = 50 $ }
\\ 
\hline
Horseshoe 
& 8.14 (1.56) & 7.96 (2.26) & 0.68 (1.35) & 10.31  (4.77)
		\\
Lasso
& 7.40 (2.09) & 9.60 (3.67) & 2.15 (1.90) & 9.83 (4.76)
		\\
		\hline
\multicolumn{5}{c }{ $ s^* = 50,  \rho_X = 0.5  $ }
\\ 
\hline
Horseshoe 
& 7.73 (2.14) & 7.18 (1.94) & 0.03 (0.02) &  7.63 (3.24)
		\\
Lasso
& 3.45 (1.19) & 6.87 (2.79) & 1.36 (0.74) & 5.95 (2.92)
		\\
		\hline
		\hline	      
\end{tabular}
\label{tb_p300}
\end{table}

\begin{table}[!h]
\centering
	\caption{Simulation results for $ p = 1000, n = 500 $.}
	\begin{tabular}{  l ccc c  }
		\hline \hline
Method  & $ 10^3 \times \ell_2(\beta_0)  $ 
& $ \ell_2( X^\top\!\! \beta_0) $ 
& $ \ell_2(Y)$ 
& $ \ell_2(Y_{\rm test}) $ 
		\\
		\hline
\multicolumn{5}{c }{ $ s^* = 10 $ }
\\ 
\hline
Horseshoe 
& 0.20 (0.04) & 0.30 (0.06) & 0.12 (0.03) & 0.53 (0.21)
		\\
Lasso
& 0.37 (0.11) & 0.33 (0.10) & 0.51 (0.09) & 0.63 (0.26)
		\\
		\hline
\multicolumn{5}{c }{ $ s^* = 10,  \rho_X = 0.5 $ }
\\ 
\hline
Horseshoe 
& 0.21 (0.04) & 0.30 (0.05) & 0.14 (0.03) & 0.57 (0.29)
		\\
Lasso 
& 0.30 (0.10) & 0.28 (0.09) & 0.53 (0.08) & 0.59 (0.28)
\\
		\hline
\multicolumn{5}{c }{ $ s^* = 50 $ }
\\ 
\hline
Horseshoe 
& 0.42 (0.07) & 0.48 (0.05) & 0.17 (0.13) & 0.64 (0.29)
		\\
Lasso
& 4.60 (1.29) & 2.99 (0.81) & 1.00 (0.39) & 2.50 (1.37)
		\\
		\hline
\multicolumn{5}{c }{ $ s^* = 50,  \rho_X = 0.5  $ }
\\ 
\hline
Horseshoe 
& 0.74 (0.22) & 0.53 (0.09) & 0.36 (0.39) &  0.75 (0.29)
		\\
Lasso
& 1.56 (0.36) & 1.51 (0.44) & 0.83 (0.45) & 1.48 (0.64)
		\\
		\hline
		\hline	
\end{tabular}
\label{tb_p1000n500}
\end{table}

Some trace plots and ACF plots for the Gibbs sampler are further given in Figure \ref{fig_tracplot}  and \ref{fig_acf_plot} in the Appendix.

\paragraph{Results with large sample sizes}

Table \ref{tb_largesample} presents the results for scenarios with large sample sizes. The simulation setup remains consistent with previous experiments, with the number of predictors fixed at $p = 100$ and the true sparsity level set to $s^* = 10$. The sample size is varied across $n \in \{200, 1000, 5000\}$, and each configuration is repeated 100 times. The results indicate that even in large-sample settings, the Horseshoe method consistently outperforms the Lasso, particularly in terms of estimation error, where it achieves approximately twice the accuracy of the Lasso.

\begin{table}[!h]
\centering
\caption{Simulation results for large sample size with $ p = 100, s^* =10 $.}
	\begin{tabular}{  l ccc c  }
		\hline \hline
  & $  \ell_2(\beta_0) $ 
& $ \ell_2( X^\top\!\! \beta_0) $ 
& $ \ell_2(Y)$ 
& $ \ell_2(Y_{\rm test}) $ 
		\\
		\hline
\multicolumn{5}{l }{ $ n = 200 $ }
\\ 
Horseshoe
& 0.23$ \times 10^{-2} $  (0.09$ \times 10^{-2} $) 
& 0.22 (0.08) & 0.27 (0.06) & 0.55 (0.23)
		\\
Lasso
& 0.48$ \times 10^{-2} $ (0.16$ \times 10^{-2} $) 
& 0.41 (0.14) & 0.44 (0.12) &  0.66 (0.28)
		\\
		\hline
\multicolumn{5}{l }{ $ n = 1000 $ }
\\ 
Horseshoe 
& 0.37$ \times 10^{-3} $ (0.12$ \times 10^{-3} $) & 0.37$ \times 10^{-1} $ (0.11$ \times 10^{-1} $) & 0.42 (0.03) & 0.50 (0.18)
		\\
Lasso 
& 0.83$ \times 10^{-3} $ (0.29$ \times 10^{-3} $) & 0.81$ \times 10^{-1} $ (0.28$ \times 10^{-1} $) & 0.46 (0.04) & 0.53 (0.20)
\\
		\hline
\multicolumn{4}{l }{ $ n = 5000 $ }
\\ 
Horseshoe 
& 0.07$ \times 10^{-3} $ (0.02$ \times 10^{-3} $) & 0.07$ \times 10^{-1} $ (0.02$ \times 10^{-1} $) & 0.46 (0.02) & 0.47 (0.18)
		\\
Lasso
& 0.16$ \times 10^{-3} $ (0.06$ \times 10^{-3} $) & 0.16$ \times 10^{-1} $ (0.06$ \times 10^{-1} $) & 0.47 (0.02) & 0.49 (0.18)
		\\
		\hline
		\hline	
\end{tabular}
\label{tb_largesample}
\end{table}

\paragraph{Results with model violation}
We now explore the robustness of our method and the Lasso to deviations from the Normal noise assumption through a brief simulation study. 
Specifically, we repeat the previous simulation setup, but with a heavier-tailed noise: the response is generated as $y_i^* = X_i^\top \beta_0 + \epsilon_i$, where $\epsilon_i \sim t_3$, a Student's t-distribution with 3 degrees of freedom. 
Table \ref{tb_p100_student} presents the results averaged over 100 repetitions. As expected, both methods exhibit slightly increased errors compared to the Normal case (Table \ref{tb_p100}), with the Horseshoe generally outperforming the Lasso.

\begin{table}[!h]
\centering
	\caption{Simulation results for model violation with $ p = 100, n = 80 $.}
	\begin{tabular}{  l ccc c  }
		\hline \hline
Method  & $ 10 \times \ell_2(\beta_0) $ 
& $ \ell_2( X^\top\!\! \beta_0) $ 
& $ \ell_2(Y)$ 
& $ \ell_2(Y_{\rm test}) $ 
		\\
		\hline
\multicolumn{5}{c }{ $ s^* = 10 $ }
\\ 
\hline
Horseshoe 
& 0.37 (0.14)  & 2.06 (0.67) & 0.54 (0.37) & 2.70 (1.39)
		\\
Lasso
& 0.40 (0.15) & 2.31 (0.86) & 0.94 (0.79) & 2.89 (1.52)
		\\
		\hline
\multicolumn{5}{c }{ $ s^* = 10,  \rho_X = 0.5 $ }
\\ 
\hline
Horseshoe 
& 0.44 (0.31) & 2.49 (3.95) & 1.80 (9.04) & 2.45 (1.56)
		\\
Lasso 
& 0.35 (0.22) &  2.38 (1.65) & 3.08 (13.6) & 2.48 (1.48)
\\
		\hline
\multicolumn{5}{c }{ $ s^* = 50 $ }
\\ 
\hline
Horseshoe 
& 3.31 (0.35) & 12.3  (3.21) & 2.60 (1.17) & 16.3  (9.43)
		\\
Lasso
& 3.72 (0.51) & 16.2 (8.46) & 4.29 (5.57) & 17.5  (9.39)
		\\
		\hline
\multicolumn{5}{c }{ $ s^* = 50,  \rho_X = 0.5  $ }
\\ 
\hline
Horseshoe 
& 2.32 (0.51) & 12.9 (3.71) & 1.77 (0.50) & 17.4  (8.22)
		\\
Lasso
& 3.27 (0.58) & 18.7 (7.28) & 3.41 (2.86) & 21.7  (9.40)
		\\
		\hline
		\hline	
\end{tabular}
\label{tb_p100_student}
\end{table}

\begin{figure}[!ht]
    \centering
    \includegraphics[width=6cm]{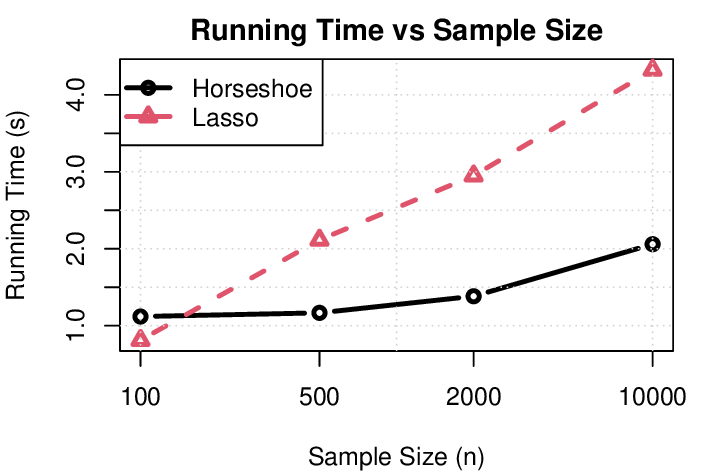}
\includegraphics[width=6cm]{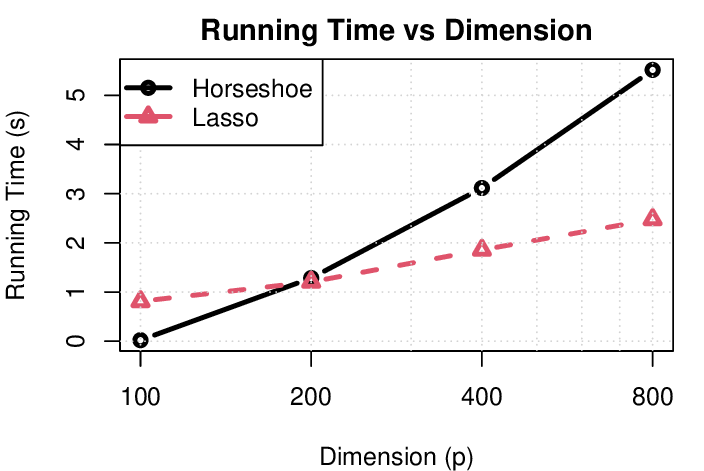}
    \caption{Log-log plot of computation times. Left: Varying sample size with fixed $p = 100, s^* = 10$. Right: Varying number of predictors with fixed $n = 100, s^* = 10$.}
    \label{fig_runingtime}
\end{figure}

\paragraph{Results on running times}
We provide a brief summary of the computation times for the Horseshoe and Lasso methods under two scenarios: increasing sample size $n$ and increasing number of predictors $p$. 
Figure \ref{fig_runingtime} shows the runtime results averaged over 10 random repetitions for each scenario. As expected, the Horseshoe method becomes computationally intensive when $p > n$, due to the need to invert a $p \times p$ matrix at each iteration. 
However, when the dimensionality is fixed and the sample size is large, the Horseshoe method proves to be significantly more efficient than Lasso in terms of computation time, and it also yields lower estimation error (see Table \ref{tb_largesample}). 
All computations were performed on a  laptop with an Intel(R) Core(TM) Ultra 7 165H processor, 68GB of RAM, running R version 4.4.0 on a 64-bit Windows 11 platform.

\subsection{Application: censored  prediction using gene expression data}

Here, we explore predictive modeling using high-dimensional genomic data originally presented by \cite{scheetz2006regulation}. Their research involved measuring RNA expression in the eyes of 120 male rats, each 12 weeks old, across 31,042 probe sets. Our interest lies in predicting the expression levels of the TRIM32 gene, which was linked to Bardet-Biedl syndrome by \cite{chiang2006homozygosity}. Given that a substantial number of probes were found to be inactive in the eye, as noted in \cite{scheetz2006regulation}, we adopt narrow our focus to the 500 genes most strongly correlated with TRIM32 expression as in \cite{huang2010variable,sherwood2022additive}. To create censored response, we additionally truncate response values below the median by replacing them with the median itself.
The dataset is publicly accessible via the `\texttt{abess}` package in \texttt{R}, as described by \cite{zhu2022abess}.

\begin{table}[!h]
	\centering
	\caption{Mean (and standard deviation) censored prediction errors for the TRIM32 data.}
	\begin{tabular}{  l | cc }
		\hline \hline
 & Horseshoe & Lasso 	
 \\ 		\hline
$ \ell_2(Y_{\rm test}) $ & 0.169 (0.053) & 0.184 (0.064)
\\
		\hline
		\hline
	\end{tabular}
	\label{tb_trim32_censored}
\end{table}

To assess the methods, we randomly allocate 84 of the 120 samples for training and the remaining 36 for testing, maintaining an approximate 70/30 percent of the data split. The methods are executed using the training set, and their prediction accuracy is evaluated on the test set. This procedure is repeated 100 times, each with a different random partition of the data. 
The outcomes of these iterations are displayed in Table \ref{tb_trim32_censored}. 
In this real dataset analysis, the Horseshoe method demonstrates superior performance compared to the Lasso, consistent with its strong results observed in the simulation studies.

\section{Conclusion}
\label{sc_conclusion}
In this work, we developed a novel Bayesian framework for high-dimensional censored regression by integrating the fractional posterior with the Tobit model and the Horseshoe prior. Our approach addresses a critical gap in the literature by combining three key elements: (i) a principled Bayesian treatment of censoring via the Tobit model, (ii) sparsity-inducing regularization through the Horseshoe prior, and (iii) robust, theoretically grounded inference via the fractional posterior. To our knowledge, this represents the first contribution to establish posterior consistency and concentration rates for Bayesian Tobit regression in high-dimensional settings.

From a methodological standpoint, we introduced a tailored Gibbs sampling algorithm based on data augmentation and the Horseshoe prior’s hierarchical structure, enabling efficient and exact Bayesian inference. Our simulation studies show that the proposed method delivers strong empirical performance, consistently outperforming recent Lasso-based approaches, especially in sparse regimes. We also demonstrated the practical value of our approach through real data analysis.

Our findings open several avenues for future research, including extensions to more complex censoring structures (e.g., interval censoring), models with non-Gaussian error distributions, and scalable variational inference methods \cite{johndrow2020scalable}. In addition, establishing theoretical guarantees for the standard posterior case (i.e., when $\alpha = 1$) remains an important direction for future investigation.

\subsection*{Acknowledgments}
The views, results, and opinions presented in this paper are solely those of the author and do not, in any form, represent those of the Norwegian Institute of Public Health.

\subsection*{Conflicts of interest/Competing interests}
The author declares no potential conflict of interests.

\clearpage
\appendix

\section{Proofs}
\label{sc_proofs}

\begin{proof}[\bf Proof of Theorem~\ref{theorem_result_dis_expectation}]
We will first apply Theorem 2.6 in \cite{alquier2020concentration} to obtain our consistency result. In order to do so, we need to check the conditions given in Theorem \ref{thm_expect_alquier}.

From Assumption \ref{asmum_lip_alquier} we have that the log-likelihood satisfies that
	\[
\vert\log p_{\beta_0} (X)
-
\log p_{\beta}(X) \vert 
\leq 
h(X ) \Vert\beta-\beta_0 \Vert_2^2
	.
	\]
Thus, we have 
\begin{align*}
\mathcal{K}(P_{\beta_0},P_\beta)
& =
\mathbb{E}\left[\log p_{\beta_0}(X ) - \log p_\beta(X )\right]
\\
&	\leq 
\mathbb{E} h(X ) \Vert\beta-\beta_0\Vert_2^2
\\
&	\leq 
B_1 \Vert\beta-\beta_0\Vert_2^2
.    
\end{align*}
When integrating with respect to 
\begin{equation}
    \label{eq_specific_distr}
\rho_n \propto \mathbf{1}_{ \{\| \beta-\beta_0 \|_2 < \delta \} } 
\pi_{HS}
,
\end{equation}
for $\delta=\{s^*\log (p/s^*)/n\}^{1/2}$, we have that
	\begin{align*}
	\int\mathcal{K}(P_{\beta_0},P_{\beta})
    \rho_n({\rm d}\beta)
&	\leq
B_1	\int \Vert\beta-\beta_0\Vert_2^2 \,
 \rho_n({\rm d}\beta)
\\
&	\leq 
B_1 \delta^2 
= 
B_1 \frac{s^*\log (p/s^*)}{n} 
	,
	\end{align*}
From Lemma \ref{lm_bound_prior_horseshoe}, we have that 
\begin{align*}
\frac{1}{n} \mathcal{K}(\rho_n,\pi)
&	\leq
\frac{1}{n} \log \frac{1}{	\pi_{HS} (\|\beta - \beta_0 \|_2<\delta)}
\\
& \leq
	K \frac{s^*\log (p/s^*)}{n} 
	.
\end{align*}
Therefore, we can now apply Theorem \ref{thm_expect_alquier}
 with $ \rho_n $ given in \eqref{eq_specific_distr} and
$$
\varepsilon_n
=
\mathcal{C} \frac{s^*\log (p/s^*)}{n} 
,
$$
where $ \mathcal{C}>0 $ is some numerical constant depending only on $ C_1,B_1 $.
Thus, we obtain the result in \eqref{eq_consistency}.

To obtain concentration result in \eqref{eq_concentration}, 
we can check the conditions required in Theorem 2.4 in \cite{alquier2020concentration}, see Theorem \ref{thm_concentration_alquier} below. 
From Assumption \ref{asmum_lip_alquier} we have that  
\begin{align*}
\mathcal{K}(P_{\beta_0},P_\beta)
& =
\mathbb{E}\left[\log p_{\beta_0}(X ) - \log p_\beta(X )\right]
\\
&	\leq 
\mathbb{E} h(X ) \Vert\beta-\beta_0\Vert_2^2
\\
&	\leq 
B_1 \Vert\beta-\beta_0\Vert_2^2
.    
\end{align*}
and, one also obtains that
\begin{align*}
\mathbb{E}\left[\log^2\frac{p_{\beta_0}}{p_\beta}(X ) \right]
& = 
\mathbb{E}\left[\left(
\log p_{\beta_0}(X  ) - \log p_\beta(X  ) \right)^2 \right]
\\
& \leq 
	\mathbb{E}h^2(X )
	\Vert\beta-\beta_0\Vert_2^4
 \\
& \leq 
B_2	\Vert\beta-\beta_0\Vert_2^4
	.
\end{align*}
When integrating with respect to $\rho_n  $ given in \eqref{eq_specific_distr}, we have that
	\begin{align*}
\int\mathbb{E}\left[\log^2\frac{p_{\beta_0}}{p_\beta}(X)\right]\rho_n({\rm d} \beta)
&	\leq 
B_2	\int \Vert\beta-\beta_0\Vert_2^4 \,
 \rho_n({\rm d}\beta)
 \\
& \leq
B_2 \delta^4
=
B_2 \left[ \frac{s^*\log (p/s^*)}{n} \right]^2
,
	\end{align*}	
From Lemma \ref{lm_bound_prior_horseshoe}, we have that 
\begin{align*}
\frac{1}{n} \mathcal{K}(\rho_n,\pi)
&	\leq
\frac{1}{n} \log \frac{1}{	\pi_{HS} (\|\beta - \beta_0 \|_2<\delta)}
\\
& \leq
	K \frac{s^*\log (p/s^*)}{n} 
	.
\end{align*}
Putting all together, one has that
	\begin{align*}
\int\mathbb{E}\left[\log^2\frac{p_{\beta_0}}{p_\beta}(X)\right]\rho_n({\rm d} \beta)
&	\leq 
B_2 \left[ \frac{s^*\log (p/s^*)}{n} \right]^2
	,
\\
	\int\mathcal{K}(P_{\beta_0},P_{\beta})\rho_n({\rm d} \beta)
&	\leq 
B_1 \frac{s^*\log (p/s^*)}{n} 
	,
    \\
	\frac1n \mathcal{K}(\rho_n,\pi)
&	\leq
K \frac{s^*\log (p/s^*)}{n} 
	.
	\end{align*}
Consequently, we can now apply Theorem 2.4 from \cite{alquier2020concentration}, with $\rho_n $ given in \eqref{eq_specific_distr}, along  with
$$
\varepsilon_n
=
\max (B_1,B_2,K)  \frac{s^*\log (p/s^*)}{n} 
.
$$
The result in \eqref{eq_concentration} is obtained.
The proof is completed.	
    
\end{proof}

\subsection{Lemmas}

\begin{theorem}[Theorem 2.6 in \cite{alquier2020concentration}]
\label{thm_expect_alquier}
Assume that $\varepsilon_n>0$ is such that there is distribution $\rho_n $ such that
$$
  \int \mathcal{K}(P_{\beta_0},P_{\beta}) \rho_n({\rm d}\beta)  \leq\varepsilon_n
  \text{; and } \,
  \mathcal{K}(\rho_n,\pi)  \leq n\varepsilon_n.
$$
 Then, for any $\alpha\in(0,1)$,
$$
\mathbb{E} \left[ \int D_{\alpha}(P_{\beta},P_{\beta_0}) \pi_{n,\alpha}({\rm d}\beta|X ^n) \right]
\leq \frac{1+\alpha}{1-\alpha}\varepsilon_n.
$$
\end{theorem}

\begin{theorem}[Corollary 2.5 in \cite{alquier2020concentration}]
\label{thm_concentration_alquier}
 Assume that a sequence $\varepsilon_n>0$ is such that there is a distribution $\rho_n $ such that
$$
  \int \mathcal{K}(P_{\beta_0},P_{\beta}) \rho_n({\rm d}\beta) \leq \varepsilon_n 
  \text{; }
  \int \mathbb{E} \log^2 \left( \frac{p_{\beta}(X_i)}{p_{\beta_0}(X_i)} \right)   \rho_n({\rm d}\beta) \leq \varepsilon_n
 \text{; } \,
  \mathcal{K}(\rho_n,\pi) \leq n \varepsilon_n.
$$
 Then, for any $\alpha\in(0,1)$, 
  \begin{equation*}
  \mathbb{P}\left[
 \int D_{\alpha}(P_{\beta},P_{\beta_0}) \pi_{n,\alpha}({\rm d}\beta|X ^n)  \leq  \frac{2(\alpha+1)}{1-\alpha} \varepsilon_n
 \right]
 \geq 1-\frac{2}{n\varepsilon_n}
 .
 \end{equation*}
\end{theorem}

\begin{lemma}[Lemma 3 in \cite{mai2024concentration}]
	\label{lm_bound_prior_horseshoe}
	Suppose $ \beta_0 \in \mathbb{R}^p $ such that $ \|\beta_0\|_0 = s^* $ and  that $ s^* < n < p $ and $ \|\beta_0\|_\infty \leq C_1  $. Suppose $\beta \sim \pi_{HS} $. 
    Define $\delta=\{s^*\log (p/s^*)/n\}^{1/2} $. Then we have, for some constant $K>0 $, that
$$
\pi_{HS} ( 
\|\beta -\beta_0\|_2<\delta )
	\geq 
	e^{-Ks^*\log (p/s^*)}.
	$$
\end{lemma}

\clearpage
\section{Additional results for simulations}

Here, we additionally illustrate the behavior of the Gibbs sampler using the trace plot and ACF plot shown in Figures \ref{fig_tracplot} and \ref{fig_acf_plot} below. These results are based on the simulation with setting $p = 300$, $n = 200$, $s^* = 10$, and $\rho_X = 0$.

\begin{figure}[!ht]
    \centering
    \includegraphics[width=12cm]{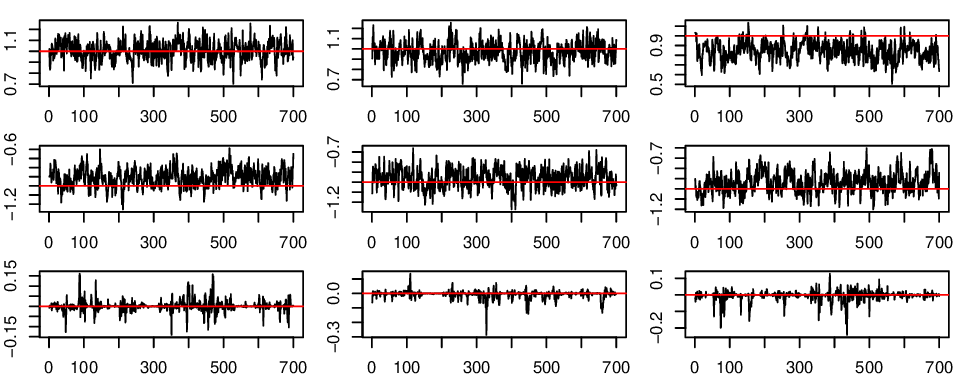}
    \caption{Trace plots from the Gibbs sampler for selected parameter entries.
Top row: three randomly chosen entries with true value 1.
Middle row: three randomly chosen entries with true value $-1$.
Bottom row: three randomly chosen entries with true value 0. Red lines show the true values. }
    \label{fig_tracplot}
\end{figure}

\begin{figure}[!ht]
    \centering
    \includegraphics[width=12cm]{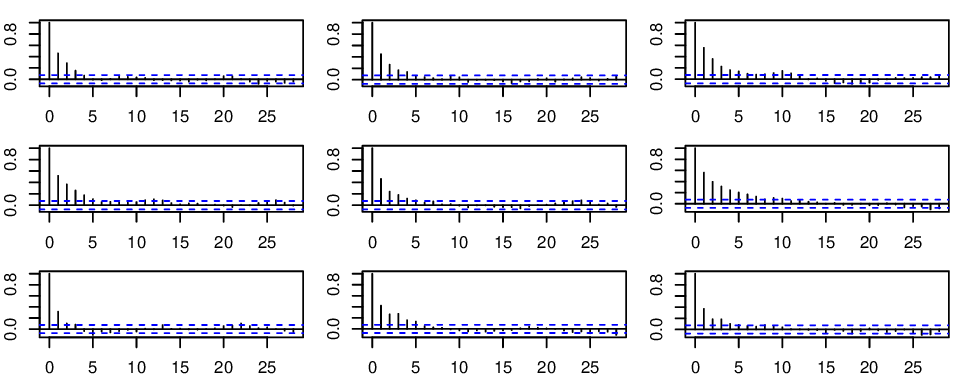}
    \caption{ACF plots from the Gibbs sampler for some random entries as in Figure \ref{fig_tracplot}. Top row (3 plots): 3 random entries with true value 1. Middle row (3 plots): 3 random entries with true value $-1$. Bottom row (3 plots): 3 random entries with true value 0.}
    \label{fig_acf_plot}
\end{figure}


\clearpage
\begin{thebibliography}{}
	
	\bibitem[Abbas, 2019]{abbas2019bayesian}
	Abbas, H.~K. (2019).
	\newblock Bayesian lasso tobit regression.
	\newblock {\em Journal of AL-Qadisiyah for computer science and mathematics},
	11(2):1--13.
	
	\bibitem[Abramovich and Grinshtein, 2016]{abramovich2016model}
	Abramovich, F. and Grinshtein, V. (2016).
	\newblock Model selection and minimax estimation in generalized linear models.
	\newblock {\em IEEE Transactions on Information Theory}, 62(6):3721--3730.
	
	\bibitem[Alhamzawi, 2020]{alhamzawi2020new}
	Alhamzawi, A. (2020).
	\newblock {A new Bayesian elastic net for tobit regression}.
	\newblock {\em Journal of Physics: Conference Series}, 1664(1):012047.
	
	\bibitem[Alhamzawi, 2016]{alhamzawi2016bayesian}
	Alhamzawi, R. (2016).
	\newblock {Bayesian elastic net Tobit quantile regression}.
	\newblock {\em Communications in Statistics-Simulation and Computation},
	45(7):2409--2427.
	
	\bibitem[Alhamzawi and Ali, 2018]{alhamzawi2018bayesian}
	Alhamzawi, R. and Ali, H. T.~M. (2018).
	\newblock Bayesian tobit quantile regression with penalty.
	\newblock {\em Communications in Statistics-Simulation and Computation},
	47(6):1739--1750.
	
	\bibitem[Alhamzawi and Yu, 2015]{alhamzawi2015bayesian}
	Alhamzawi, R. and Yu, K. (2015).
	\newblock Bayesian tobit quantile regression using g-prior distribution with
	ridge parameter.
	\newblock {\em Journal of Statistical Computation and Simulation},
	85(14):2903--2918.
	
	\bibitem[Alhusseini and Georgescu, 2018]{alhusseini2018bayesian}
	Alhusseini, F. H.~H. and Georgescu, V. (2018).
	\newblock Bayesian composite tobit quantile regression.
	\newblock {\em Journal of Applied Statistics}, 45(4):727--739.
	
	\bibitem[Alquier and Ridgway, 2020]{alquier2020concentration}
	Alquier, P. and Ridgway, J. (2020).
	\newblock Concentration of tempered posteriors and of their variational
	approximations.
	\newblock {\em The Annals of Statistics}, 48(3):1475--1497.
	
	\bibitem[Amemiya, 1984]{amemiya1984tobit}
	Amemiya, T. (1984).
	\newblock Tobit models: A survey.
	\newblock {\em Journal of econometrics}, 24(1-2):3--61.
	
	\bibitem[Anceschi et~al., 2023]{anceschi2023bayesian}
	Anceschi, N., Fasano, A., Durante, D., and Zanella, G. (2023).
	\newblock Bayesian conjugacy in probit, tobit, multinomial probit and
	extensions: A review and new results.
	\newblock {\em Journal of the American Statistical Association},
	118(542):1451--1469.
	
	\bibitem[Basson et~al., 2023]{basson2023variational}
	Basson, M., Louw, T.~M., and Smith, T.~R. (2023).
	\newblock Variational tobit gaussian process regression.
	\newblock {\em Statistics and Computing}, 33(3):64.
	
	\bibitem[Bellec et~al., 2018]{bellec2018slope}
	Bellec, P.~C., Lecu{\'e}, G., and Tsybakov, A.~B. (2018).
	\newblock Slope meets lasso: improved oracle bounds and optimality.
	\newblock {\em The Annals of Statistics}, 46(6B):3603--3642.
	
	\bibitem[Bhadra et~al., 2017]{bhadra2015horseshoe_}
	Bhadra, A., Datta, J., Polson, N.~G., and Willard, B. (2017).
	\newblock The {{Horseshoe}}+ {{Estimator}} of {{Ultra}}-{{Sparse Signals}}.
	\newblock {\em Bayesian Analysis}, 12(4):1105--1131.
	
	\bibitem[Bhattacharya et~al., 2019]{bhattacharya2016bayesian}
	Bhattacharya, A., Pati, D., and Yang, Y. (2019).
	\newblock Bayesian fractional posteriors.
	\newblock {\em Annals of Statistics}, 47(1):39--66.
	
	\bibitem[Bissiri et~al., 2016]{bissiri2013general}
	Bissiri, P.~G., Holmes, C.~C., and Walker, S.~G. (2016).
	\newblock A general framework for updating belief distributions.
	\newblock {\em Journal of the Royal Statistical Society Series B: Statistical
		Methodology}, 78(5):1103--1130.
	
	\bibitem[Buckley and James, 1979]{buckley1979linear}
	Buckley, J. and James, I. (1979).
	\newblock Linear regression with censored data.
	\newblock {\em Biometrika}, 66(3):429--436.
	
	\bibitem[Carvalho et~al., 2010]{carvalho2010horseshoe}
	Carvalho, C.~M., Polson, N.~G., and Scott, J.~G. (2010).
	\newblock The horseshoe estimator for sparse signals.
	\newblock {\em Biometrika}, 97(2):465--480.
	
	\bibitem[Castillo et~al., 2015]{castillo2015bayesian}
	Castillo, I., Schmidt-Hieber, J., and {van der Vaart}, A. (2015).
	\newblock Bayesian linear regression with sparse priors.
	\newblock {\em The Annals of Statistics}, 43(5):1986--2018.
	
	\bibitem[Chiang et~al., 2006]{chiang2006homozygosity}
	Chiang, A.~P., Beck, J.~S., Yen, H.-J., Tayeh, M.~K., Scheetz, T.~E.,
	Swiderski, R.~E., Nishimura, D.~Y., Braun, T.~A., Kim, K.-Y.~A., Huang, J.,
	et~al. (2006).
	\newblock {Homozygosity mapping with SNP arrays identifies TRIM32, an E3
		ubiquitin ligase, as a Bardet--Biedl syndrome gene (BBS11)}.
	\newblock {\em Proceedings of the National Academy of Sciences},
	103(16):6287--6292.
	
	\bibitem[Chib, 1992]{chib1992bayes}
	Chib, S. (1992).
	\newblock {Bayes inference in the Tobit censored regression model}.
	\newblock {\em Journal of Econometrics}, 51(1-2):79--99.
	
	\bibitem[Gandhi et~al., 2020]{gandhi2020long}
	Gandhi, R.~T., Tashima, K.~T., Smeaton, L.~M., Vu, V., Ritz, J., Andrade, A.,
	Eron, J.~J., Hogg, E., and Fichtenbaum, C.~J. (2020).
	\newblock Long-term outcomes in a large randomized trial of hiv-1 salvage
	therapy: 96-week results of aids clinical trials group a5241 (options).
	\newblock {\em The Journal of Infectious Diseases}, 221(9):1407--1415.
	
	\bibitem[Helsel, 2011]{helsel2011statistics}
	Helsel, D.~R. (2011).
	\newblock {\em Statistics for censored environmental data using Minitab and R},
	volume~77.
	\newblock John Wiley \& Sons.
	
	\bibitem[Huang et~al., 2010]{huang2010variable}
	Huang, J., Horowitz, J.~L., and Wei, F. (2010).
	\newblock Variable selection in nonparametric additive models.
	\newblock {\em Annals of statistics}, 38(4):2282.
	
	\bibitem[Jacobson and Zou, 2024]{jacobson2024high}
	Jacobson, T. and Zou, H. (2024).
	\newblock {High-dimensional censored regression via the penalized Tobit
		likelihood}.
	\newblock {\em Journal of Business \& Economic Statistics}, 42(1):286--297.
	
	\bibitem[Jeong and Ghosal, 2021]{jeong2021posterior}
	Jeong, S. and Ghosal, S. (2021).
	\newblock Posterior contraction in sparse generalized linear models.
	\newblock {\em Biometrika}, 108(2):367--379.
	
	\bibitem[Ji et~al., 2012]{ji2012model}
	Ji, Y., Lin, N., and Zhang, B. (2012).
	\newblock Model selection in binary and tobit quantile regression using the
	gibbs sampler.
	\newblock {\em Computational Statistics \& Data Analysis}, 56(4):827--839.
	
	\bibitem[Johndrow et~al., 2020]{johndrow2020scalable}
	Johndrow, J., Orenstein, P., and Bhattacharya, A. (2020).
	\newblock {Scalable approximate MCMC algorithms for the horseshoe prior}.
	\newblock {\em Journal of Machine Learning Research}, 21(73):1--61.
	
	\bibitem[Johnson, 2009]{johnson2009lasso}
	Johnson, B.~A. (2009).
	\newblock On lasso for censored data.
	\newblock {\em Electronic Journal of Statistics}, 3:485--506.
	
	\bibitem[Knoblauch et~al., 2022]{Knoblauch}
	Knoblauch, J., Jewson, J., and Damoulas, T. (2022).
	\newblock {An Optimization-centric View on Bayes' Rule: Reviewing and
		Generalizing Variational Inference}.
	\newblock {\em Journal of Machine Learning Research}, 23(132):1--109.
	
	\bibitem[Kobayashi, 2017]{kobayashi2017bayesian}
	Kobayashi, G. (2017).
	\newblock Bayesian endogenous tobit quantile regression.
	\newblock {\em Bayesian Analysis}, 12(1):161--191.
	
	\bibitem[Li et~al., 2014]{li2014dantzig}
	Li, Y., Dicker, L., and Zhao, S.~D. (2014).
	\newblock {The Dantzig selector for censored linear regression models}.
	\newblock {\em Statistica Sinica}, 24(1):251.
	
	\bibitem[Mai, 2024a]{mai2024concentration}
	Mai, T.~T. (2024a).
	\newblock {Concentration of a Sparse Bayesian Model With Horseshoe Prior in
		Estimating High-Dimensional Precision Matrix}.
	\newblock {\em Stat}, 13(4):e70008.
	
	\bibitem[Mai, 2024b]{mai2024gbayslogistic}
	Mai, T.~T. (2024b).
	\newblock On high-dimensional classification by sparse generalized bayesian
	logistic regression.
	\newblock {\em arXiv preprint arXiv:2403.12832}.
	
	\bibitem[Mai, 2024c]{mai2024properties}
	Mai, T.~T. (2024c).
	\newblock On properties of fractional posterior in generalized reduced-rank
	regression.
	\newblock {\em arXiv preprint arXiv:2404.17850}.
	
	\bibitem[Mai, 2025]{mai2025concentration}
	Mai, T.~T. (2025).
	\newblock Concentration properties of fractional posterior in 1-bit matrix
	completion.
	\newblock {\em Machine Learning}, 114(1):7.
	
	\bibitem[Makalic and Schmidt, 2015]{makalic2015simple}
	Makalic, E. and Schmidt, D.~F. (2015).
	\newblock {A Simple Sampler for the Horseshoe Estimator}.
	\newblock {\em IEEE Signal Processing Letters}, 23(1):179--182.
	
	\bibitem[Martin et~al., 2017]{martin2017empirical}
	Martin, R., Mess, R., and Walker, S.~G. (2017).
	\newblock {Empirical Bayes posterior concentration in sparse high-dimensional
		linear models}.
	\newblock {\em Bernoulli}, 23(3):1822--1847.
	
	\bibitem[Martin and Tang, 2020]{martin2020empirical}
	Martin, R. and Tang, Y. (2020).
	\newblock Empirical priors for prediction in sparse high-dimensional linear
	regression.
	\newblock {\em Journal of Machine Learning Research}, 21(144):1--30.
	
	\bibitem[McDonald and Moffitt, 1980]{mcdonald1980uses}
	McDonald, J.~F. and Moffitt, R.~A. (1980).
	\newblock {The uses of Tobit analysis}.
	\newblock {\em The review of economics and statistics}, pages 318--321.
	
	\bibitem[Meeker et~al., 2021]{meeker2021statistical}
	Meeker, W.~Q., Escobar, L.~A., and Pascual, F.~G. (2021).
	\newblock {\em Statistical methods for reliability data}.
	\newblock John Wiley \& Sons.
	
	\bibitem[M{\"u}ller and van~de Geer, 2016]{muller2016censored}
	M{\"u}ller, P. and van~de Geer, S. (2016).
	\newblock Censored linear model in high dimensions: Penalised linear regression
	on high-dimensional data with left-censored response variable.
	\newblock {\em Test}, 25:75--92.
	
	\bibitem[O’Neill, 2024]{o2024type}
	O’Neill, E. (2024).
	\newblock {Type I tobit Bayesian additive regression trees for censored outcome
		regression}.
	\newblock {\em Statistics and Computing}, 34(4):123.
	
	\bibitem[Piironen and Vehtari, 2017]{piironen2017sparsity}
	Piironen, J. and Vehtari, A. (2017).
	\newblock Sparsity information and regularization in the horseshoe and other
	shrinkage priors.
	\newblock {\em Electronic Journal of Statistics}, 11(2):5018--5051.
	
	\bibitem[Powell, 1984]{powell1984least}
	Powell, J.~L. (1984).
	\newblock Least absolute deviations estimation for the censored regression
	model.
	\newblock {\em Journal of econometrics}, 25(3):303--325.
	
	\bibitem[Rigollet, 2012]{rigollet2012kullback}
	Rigollet, P. (2012).
	\newblock Kullback-leibler aggregation and misspecified generalized linear
	models.
	\newblock {\em The Annals of Statistics}, 40(2):639--665.
	
	\bibitem[Scheetz et~al., 2006]{scheetz2006regulation}
	Scheetz, T.~E., Kim, K.-Y.~A., Swiderski, R.~E., Philp, A.~R., Braun, T.~A.,
	Knudtson, K.~L., Dorrance, A.~M., DiBona, G.~F., Huang, J., Casavant, T.~L.,
	et~al. (2006).
	\newblock Regulation of gene expression in the mammalian eye and its relevance
	to eye disease.
	\newblock {\em Proceedings of the National Academy of Sciences},
	103(39):14429--14434.
	
	\bibitem[Sherwood and Maidman, 2022]{sherwood2022additive}
	Sherwood, B. and Maidman, A. (2022).
	\newblock Additive nonlinear quantile regression in ultra-high dimension.
	\newblock {\em Journal of Machine Learning Research}, 23(63):1--47.
	
	\bibitem[Soret et~al., 2018]{soret2018lasso}
	Soret, P., Avalos, M., Wittkop, L., Commenges, D., and Thi{\'e}baut, R. (2018).
	\newblock Lasso regularization for left-censored gaussian outcome and
	high-dimensional predictors.
	\newblock {\em BMC medical research methodology}, 18:1--13.
	
	\bibitem[Tobin, 1958]{tobin1958estimation}
	Tobin, J. (1958).
	\newblock Estimation of relationships for limited dependent variables.
	\newblock {\em Econometrica: journal of the Econometric Society}, pages 24--36.
	
	\bibitem[van~de Geer, 2008]{van2008high}
	van~de Geer, S.~A. (2008).
	\newblock High-dimensional generalized linear models and the lasso.
	\newblock {\em The Annals of Statistics}, 36(2):614--645.
	
	\bibitem[Van~Erven and Harremos, 2014]{van2014renyi}
	Van~Erven, T. and Harremos, P. (2014).
	\newblock {R{\'e}nyi divergence and Kullback-Leibler divergence}.
	\newblock {\em IEEE Transactions on Information Theory}, 60(7):3797--3820.
	
	\bibitem[Yu and Stander, 2007]{yu2007bayesian}
	Yu, K. and Stander, J. (2007).
	\newblock Bayesian analysis of a tobit quantile regression model.
	\newblock {\em Journal of Econometrics}, 137(1):260--276.
	
	\bibitem[Zhao and Lian, 2015]{zhao2015bayesian}
	Zhao, K. and Lian, H. (2015).
	\newblock Bayesian tobit quantile regression with single-index models.
	\newblock {\em Journal of Statistical Computation and Simulation},
	85(6):1247--1263.
	
	\bibitem[Zhou and Liu, 2016]{zhou2016lad}
	Zhou, X. and Liu, G. (2016).
	\newblock Lad-lasso variable selection for doubly censored median regression
	models.
	\newblock {\em Communications in Statistics-Theory and Methods},
	45(12):3658--3667.
	
	\bibitem[Zhu et~al., 2022]{zhu2022abess}
	Zhu, J., Wang, X., Hu, L., Huang, J., Jiang, K., Zhang, Y., Lin, S., and Zhu,
	J. (2022).
	\newblock {abess: a fast best-subset selection library in python and R}.
	\newblock {\em Journal of Machine Learning Research}, 23(202):1--7.
	
\end{thebibliography}
\end{document}